\documentclass{article}

\usepackage{arxiv}

\usepackage[utf8]{inputenc} 
\usepackage[T1]{fontenc}    
\usepackage{hyperref}       
\usepackage{url}            
\usepackage{booktabs}       
\usepackage{amsfonts}       
\usepackage{nicefrac}       
\usepackage{microtype}      
\usepackage{graphicx}
\usepackage{apacite}
\usepackage{natbib}
\usepackage{doi}

\usepackage[table]{xcolor}
\usepackage{amsmath}
\newcommand{\probP}{\text{I\kern-0.15em P}}

\usepackage{xcolor}
\usepackage[dvipsnames]{xcolor}

\usepackage{eqparbox}
\newcommand{\swl}[2][nmbr]{\eqmakebox[#1]{\strut #2}}

\title{A generic nonparametric value-at-risk estimator for high dimensions}

\author{ \href{https://orcid.org/0000-0001-8802-7184}{\includegraphics[scale=0.06]{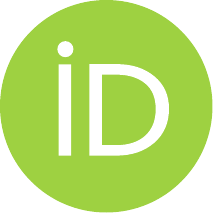}\hspace{1mm}Siyuan Sun}\\
	Investivity SA.\\
	Place de Saint-Gervais 1,\\
	Geneva 1201, Switzerland \\
	\texttt{siyuan.sun@investivity.com} \\
}

\renewcommand{\shorttitle}{Nonparametric Value-at-Risk for High Dimensions}

\hypersetup{
pdftitle={A generic nonparametric value-at-risk estimator for high dimensions},
pdfsubject={Quantitative Finance, Risk Management},
pdfauthor={Siyuan Sun},
pdfkeywords={Value at Risk, VaR, Expected Shortfall, CVaR, High-Dimensions, Non-parametric},
}

\begin{document}
\maketitle

\begin{abstract}
	We present in this article a non-parametric value-at-risk (VaR+CVaR) algorithm that remains accurate for an arbitrarily large number of underlying positions.  The algorithm solves the two inherent problems of VaR estimation.  First, past history is not directly applicable to the future, but all predictions of the future are based on the past.  Second, VaR estimation is equivalent to modeling a single corner of a high-dimensional space (the corner where all bets lose simultaneously).  The algorithm only uses mathematical methods that strictly do not degrade in accuracy at high-dimensions.  Historical data are then directly incorporated with all high-dimensional relationships present, without manipulation. We test the algorithm with an ensemble of 500 portfolios with random positions across 49 distinct liquid futures of different expiries (VIX, equity indexes, gov. bonds, rates, energy, metals, livestock, agriculture, and softs).  All VaR estimations are performed strictly blind to the future.  The median portfolio rate of loss exceeding the 99\% confidence daily VaR estimate is between $1.0\pm0.1$\% depending on algorithm input parameters. 68\% of portfolios have a rate of loss exceeding 99\% VaR between $1.0\pm0.3$\%, and 95\% of portfolios between $1.0\pm0.5$\%. 
\end{abstract}


\section{Introduction}

We present in this article a computationally simple, generic value-at-risk (VaR and CVaR) algorithm that remains accurate for an arbitrarily large number of different underlying positions. \\
~\\
The algorithm tackles the two fundamental issues inherent to all tail risk estimation:
\begin{itemize}
    \item Long-term risk is inherently not the same as short-term risk, but we need long-term history (which is not directly applicable) to say anything about future risk
    \item Tail-risk describes the case when all our bets go against us at the same time.  Once we have 100+ different positions, this is equivalent to modeling a single corner of a high-dimensional space, which is mathematically difficult.
\end{itemize}
The algorithm breaks down the probability density function of the total portfolio return into 3 constituent parts:
\begin{itemize}
    \item $expo_{k,i-1}$: the exposure to financial instrument $k$ as measured by market value / AUM at the close of day $i-1$. 
    \item $\sigma^{14}_{k,i-1}$: the recent volatility (root-mean-squared of the daily true range percentage) of $k$ in the most recent (configurable) 14 days before day $i$.
    \item $\frac{r_{k,j}}{\sigma^{14}_{k,j-1}}$: The size of $k$'s return on each past day $j$ normalized by the recent volatility before $j$.
\end{itemize}  
The recent volatility $\sigma^{14}_{k,i-1}$ adapts quickly to current situations, while the historical ratio $r_{k,j}/\sigma^{14}_{k,j-1}$ captures the non-Gaussian skew within each instrument and the relationships between all instruments $k$ in every past day $j$. The set of historical days $j$ can be selected by either a configurable look-back window or by using all available past history before day $i$. \\
\\
We will see in Section \ref{sec:2DVisualization} that the algorithm is mathematically equivalent to Monte-Carlo integration above and below a $K-1$ dimensional surface in the $K$ dimensional $r_{k,j}/\sigma^{14}_{k,j-1}$ space, where each past day $j$ is a Monte Carlo trial.  The shape of the $K-1$ dimensional surface changes according to the recent volatilities $\sigma^{14}_{k,i-1}$ and the amount of exposure $expo_{k,i-1}$ for each instrument $k$.  The higher the exposure and the higher the recent volatility, the more important the direction $k$. \\
\\
Monte-Carlo integration's error is inversely proportional to $\sqrt{N~trials}$, and performs equally well regardless of the number of total dimensions.  As such, we effectively transform the VaR estimation problem from a high-dimensional modeling problem to a ``do we have enough past data" problem.  We would argue that the ``do we have enough past data" problem always exists because all models can only be verified using past data. \\
\\
Many Monte Carlo methods in literature use models and simulations to generate trials of the rare events featured in VaR and CVaR \cite{HongMCSummary}.  For example, \cite{BouyéCopula} parameterizes the relationships between specific asset classes using copulas.  Another approach \cite{bardou2010computationvarcvarusing} uses stochastic approximations to tackle both the computational efficiency and non-linearity of rare events. Common concerns include computation optimization and using importance sampling to target the rare-events featured in tail-risk estimation \cite{GlassermanIS}. \\
\\
We contend that all simulations can only be verified using historical data. Therefore, no model or simulation can contain more information (at least, verifiable information) than that exist in past historical data.  Inevitably, simulations, modeling, and approximations either ignore part of the information in historical data and add parts that are not present.  Our approach directly extract information from unperturbed high-dimensional historical data, thereby cutting out the middleman. \\
\\
Many risk estimation algorithms and capital allocation algorithms (that depend on risk estimation) employ mathematical methods that cannot directly process high dimensions without loss of accuracy and/or stability; examples include covariance matrix inversion \cite{MeanVariance} and kernel smoothing \cite{HongKernel}.  Algorithms such as Hierarchical Risk Parity \cite{LopezHRP} and others use clustering, principal component decomposition, or other decompositions to limit the number of dimensions processed at once in order to generate a stable solution for high dimensions. The fundamental issue is that the mathematical methods used cannot be directly applied to high-dimensional situations.  Thereby, the algorithm must iteratively process low-dimensional pieces one at a time.  The iterative process often explicitly or implicitly ignores some relationships in the high-dimensional space.
\\
\\
We limit ourselves to using math that strictly does not lose accuracy with increasing dimensionality.  As such, we can directly process high-dimensional data with all its complexity, without compression or approximation. The historical ratio $r_{k,j}/\sigma^{14}_{k,j-1}$ contains all historical relationships between all instruments $k$ in every past day $j$.  Indeed, how each instrument $k$ performed relative to one another on the same past day $j$ is what we colloquially call ``correlation".  For example, the covariance matrix compresses these complex relationships down to a single box of numbers.  Instead of modeling, parameterization, and compression of these high-dimensional relationships, we take the unperturbed historical values directly as inputs. \\
\\
In Sections \ref{sec:EssembleTesting} and \ref{sec:EssembleTestingResults}, we give a blind-to-the-future back test of the VaR algorithm for an ensemble of 500 portfolios with random positions across 49 different futures of different expirations, in VIX, equity indexes, government bonds, interest rates, and the main liquid commodities (energy, livestock, agriculture, softs). We test 9 different variations of the algorithm's input parameters to confirm the stability of results for a wide range of inputs.  \\
\\
By definition, 1 percent of all days should experience a loss that exceeds the 99\% daily VaR. This translates to roughly 2 to 3 days per year and should be directly observable. \\
\\
95\% of the 500 portfolios have between a 0.5\% to 1.5\% rate of loss exceeding the 99\% VaR estimate.  68\% of all portfolios have a 0.7\% to 1.3\% rate of loss exceeding the 99\% VaR estimate. The median portfolio has a $1.0\pm0.1$\% rate of loss exceeding the 99\% VaR estimate, depending on the algorithm's input parameters. Therefore, the VaR estimation is accurate and able to correctly predict the amount of loss expected in the worst 1 percent of days for a multi-asset portfolio with a high number of underlying positions. \\
\\
We only test estimates of the 99\% confidence daily VaR in this article.  However, the algorithm can compute VaR or CVaR at any confidence, as the algorithm computes the total portfolio return PDF.  We leave back-testing of CVaR estimates as a future exercise.  Weekly, bi-weekly, or monthly VaR variations of the algorithm are also possible with modifications, but again are beyond the scope of this article.
\section{The Probability Density Function (PDF) of Tomorrow's Portfolio Return}
\label{sec:PDF}
We begin with the time series of daily returns for each underlying financial instrument $k$ as shown in \ref{eqn:ReturnTimeSeries}. We are currently at the end of day $i-1$.  $r_{k,i}$ is the unknown return of $k$ for tomorrow, day $i$. \\
\begin{equation}
    \centering
    ...~...~...~r_{k,j-14}~...~...~...~r_{k,j-1}~r_{k,j}~...~...~...~r_{k,i-14}~...~...~...~r_{k,i-1}~[r_{k,i}]
    \label{eqn:ReturnTimeSeries}
\end{equation}
\\
We can also compute the time series of the daily true range percentage (TRP) of the instrument $k$ in \ref{eqn:percentTrueRangeTimeSeries} using Equation \ref{eqn:percentTrueRange}.
\begin{equation}
    \centering
    TRP_i = \frac{Max(high_{k,i}, close_{k,i-1}) - Min(low_{k,i}, close_{k,i-1})}{close_{k,i-1}}
    \label{eqn:percentTrueRange}
\end{equation}
\begin{equation}
    \centering
    ...~...~...~TRP_{k,j-14}~...~...~...~TRP_{k,j-1}~TRP_{k,j}~...~...~...~TRP_{k,i-14}~...~...~...~TRP_{k,i-1}~[TRP_{k,i}]
    \label{eqn:percentTrueRangeTimeSeries}
\end{equation}
\\
We are currently at the end of day $i-1$.  $TRP_{k,i}$ is unknown. However, we can compute the recent volatility defined as the root-mean-squared (RMS) of the $TRP$ in the 14 days immediately preceding $i$ from $i-1$ to $i-14$ as shown in Equation \ref{eqn:recentVolatility}. \\
\begin{equation}
    \centering
    \sigma^{14}_{k,i-1} = \sqrt{\frac{1}{14}\sum_{l=i-14}^{i-1}{TRP_{k,l}*TRP_{k,l}}}
    \label{eqn:recentVolatility}
\end{equation}
\\
We chose $TRP$ as a measure of recent volatility because the daily $TRP$ incorporates some intraday information without much additional data processing.  The overall performance is similar for other recent volatility measures, such as the root-mean-squared of the close-to-close daily returns. \\
\\
Note that each past day $j$ also has its own past recent volatility $\sigma^{14}_{k,j-1}$, calculated using days $j-1$ to $j-14$.  The set of historical days $j$ considered can be selected by either a configurable look-back window or by using all available past history before day $i$.\\
\\
The length of the look-back window and the number of days considered in the recent volatility are the only two input parameters manually set by the user.  Performance vs variations of both parameters will be tested in Section \ref{sec:EssembleTesting} - \ref{sec:EssembleTestingResults}. \\
\\
We define the PDF of the portfolio's tomorrow's future return in Equation \ref{eqn:portfolioTomorrowReturnPDF}. \\
\begin{equation}
    \centering
    \probP(portfolio~return_i) = \sum_k expo_{k,i-1} * \sigma^{14}_{k,i-1} * \frac{r_{k,j}}{\sigma^{14}_{k,j-1}} 
    \label{eqn:portfolioTomorrowReturnPDF}
\end{equation}
\\
The PDF of the portfolio's future return is a combination of 3 factors: \\
\begin{itemize}
    \item $expo_{k,i-1}$: the exposure to financial instrument $k$ as measured by market value / AUM at the close of day $i-1$. 
    \item $\sigma^{14}_{k,i-1}$: the recent volatility of $k$ in the most recent 14 days before day $i$.
    \item $\frac{r_{k,j}}{\sigma^{14}_{k,j-1}}$: how much $k$'s return on each past day $j$ deviated from the recent volatility before day $j$.
\end{itemize}
The recent volatility $\sigma^{14}_{k,i-1}$ component rapidly adapts to the current situation.  The historical $r_{k,j}/\sigma^{14}_{k,j-1}$ component contains all the non-Gaussian behaviors (down-side vs up-side skew, fat-tail, etc.) for each instrument $k$ and the relationships between all instruments $k$ observed in past historical days $j$.\\
\\
We can then calculate the VaR at any confidence because we calculated the entire portfolio's total return PDF.  The 99\% VaR is defined as the worst 1\% of the PDF. The 95\% VaR is the worst 5\% of the PDF. We can also calculate the expected short-fall as the expected value of the tail of the PDF. 

\section{Two Dimensional Visualization of the PDF}
\label{sec:2DVisualization}
\begin{figure}[ht!]
    \centering
    \includegraphics[width=0.5\linewidth]{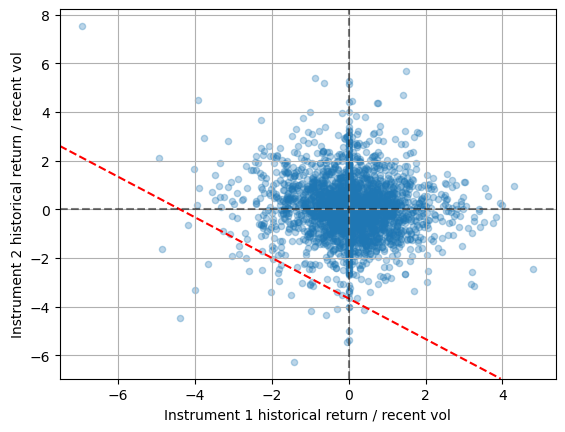}
    \caption{2D Visualization of the VaR estimation Equation \ref{eqn:VaREqn2D}.}
    \label{fig:2DCorrelationExample}
\end{figure}

\begin{figure}[ht!]
    \centering
    \includegraphics[width=0.5\linewidth]{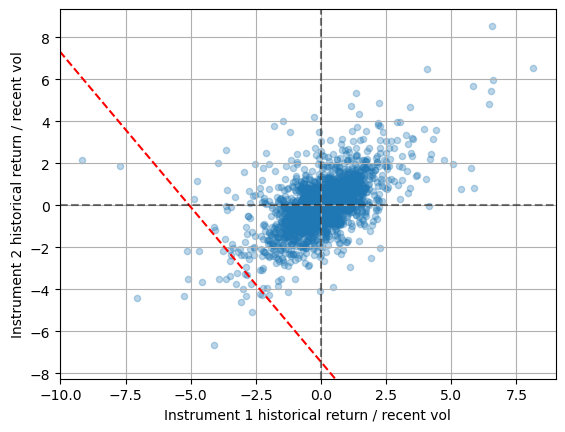}
    \caption{2D Visualization of Equation \ref{eqn:VaREqn2D} for two correlated instruments.}
    \label{fig:Rate2DCorrelationExample}
\end{figure}

We can write out Equation \ref{eqn:portfolioTomorrowReturnPDF} for instruments 1 and 2 in Equation \ref{eqn:VaREqn2D}.
\begin{equation}
    \probP(portfolio~return_i) = expo_{1,i-1} * \sigma^{14}_{1,i-1} * \frac{r_{1,j}}{\sigma^{14}_{1,j-1}} + 
          expo_{2,i-1} * \sigma^{14}_{2,i-1} * \frac{r_{2,j}}{\sigma^{14}_{2,j-1}} 
    \label{eqn:VaREqn2D}
\end{equation}
\\
Likewise, we can also plot the historical days $j$'s $r_{k,j}/\sigma^{14}_{k,j-1}$ for the 2 instruments in Figure \ref{fig:2DCorrelationExample}. \\
\\
Each historical day $j$ forms a dot with its own $(\frac{r_{1,j}}{\sigma^{14}_{1,j-1}},  \frac{r_{2,j}}{\sigma^{14}_{2,j-1}})$ as the $X$ and $Y$ coordinates. We can then substitute $X$ and $Y$ back into Equation \ref{eqn:VaREqn2D} to get Equation \ref{eqn:VaREqn2DXY}. \\
\begin{equation}
    portfolio~return_i = expo_{1,i-1} * \sigma^{14}_{1,i-1} * X + 
          expo_{2,i-1} * \sigma^{14}_{2,i-1} * Y
    \label{eqn:VaREqn2DXY}
\end{equation}
\\
Rearranging to point slope form, we get the slope for the red-dashed line, the line of equal total portfolio return, in Equation \ref{eqn:VaREqn2DXYSlope}.  The portfolio as a whole has the same total return if tomorrow's currently unknown $(X,Y)$ coordinates on day $i$ falls anywhere along the line. \\
\\
The slope of the line is completely determined by the recent volatility and the current exposure of the two instruments and changes based on the current conditions at $i-1$. The higher the recent volatility and exposure of an instrument, the more important the instrument's direction becomes.  \\
\begin{equation}
    slope = -\frac{expo_{1,i-1} * \sigma^{14}_{1,i-1}}{expo_{2,i-1}*\sigma^{14}_{2,i-1}}
    \label{eqn:VaREqn2DXYSlope}
\end{equation}
\\
The only degree of freedom remaining for the red-dashed line is the distance from the origin.  The 99\% daily VaR corresponds to the worst 1\% possibility of tomorrow's total portfolio return. In other words, we estimate tomorrow's 99\% daily VaR by how far down the line needs to move before only 1 percent of past historical days lie below the line, and 99 percent lie above the line. All else equal, the further down, the more risky tomorrow is for the portfolio as a whole. \\
\\
Notice that the algorithm organically captures the relationship between the instruments.  If two instruments are correlated, they would form a distribution such as in Figure \ref{fig:Rate2DCorrelationExample}. Here, the red-dashed line must move further down from the origin before only 1 percent of past days are below the line when compared to the case when the instruments are anti-correlated (a mirror image across the $X$ axis of Figure \ref{fig:Rate2DCorrelationExample}). A visualization of a long/short case of two correlated instruments (equivalent to being long two anti-correlated instruments) can be found in Figure \ref{fig:Cattle2DCorrelationExample}. \\
\\
If the two instruments are mostly anti-correlated, but there exist days in history that are correlated, then it would be as if the figure was a mirror image across the $X$ axis, but some dots exist in the negative X and negative Y quadrant.  The mere presence of the points in the quadrant would necessarily pull down the red-dashed line.  \\
\\
Most parameterizations (copulas, correlation matrix, kernel smears) are different ways of fitting or approximating these dots.  Approximations either add something that does not exist or ignore features that do exist in past data.  For example, the covariance, is compressing the information in this distribution to a single number. Here, we let past data speak for itself with no assumptions or approximations. \\
\\
Next, consider the process of moving the line down until only 1 percent of the past days (the dots) are below the line.  Mathematically, this procedure is Monte-Carlo integration using actual historical days $j$ as the Monte-Carlo trials.  This procedure is also known as bootstrapping using historical data. \\
\\
Modeling VaR is inherently attempting to model a single corner of a high-dimensional space, the corner where all our bets turn against us at the same time.  Analytical modeling of a single corner of a 100-dimensional space is known to be extremely difficult, if not impossible. \\
\\
The algorithm solves this by using Monte-Carlo integration of past data taken from a configurable look-back window, or from all available history for which all instruments $k$ have trading data. Monte-Carlo integration's error is inversely proportional to $\sqrt{N~trials}$ regardless of the number of dimensions and performs equally well as a computation method regardless of the number of underlying instruments $K$.\\
\\
At high $K$ dimensions, the red-dashed line generalizes to an $K-1$ surface that is orthogonal to the $<expo_{k,i-1} * \sigma^{14}_{k,i-1}>$ vector. The shape of the $K-1$ dimension surface changes according to the recent volatility and the current portfolio exposures. The only degree of freedom left, the distance from the origin, is fixed by the Monte-Carlo integration of past days $j$ on either side of the surface.  \\
\\
How each instrument performed relative to one another on the past day $j$ in the high-dimensional $r_{k,j}/\sigma^{14}_{k,j-1}$ space is what we colloquially call the correlations between all $K$ instruments. Models attempt to parameterize how each instrument's daily returns relate to one another. Instead, we allow the instruments' high-dimensional behavior in the past days $j$ to express themselves with no assumptions. The algorithm then organically picks up on these relationships using Monte-Carlo integration which be directly applied to high dimensions with no loss of accuracy. \\
\\
We essentially swap a high-dimensional modeling problem for a ``is there sufficient past data $j$ for the Monte-Carlo integration to be statistically significant" problem.  We would argue that the ``is there sufficient past data" problem always remains because models can only be verified using past data. 
\section{Relationship to Scenario-Based Stress-Testing Algorithms}
\label{sec:scenarioTesting}
\begin{figure}[ht!]
    \centering
    \includegraphics[width=0.5\linewidth]{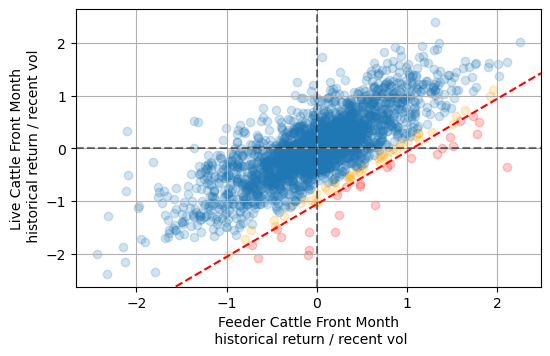}
    \caption{2D Visualization of Equation \ref{eqn:VaREqn2D} for live cattle and feeder cattle front month futures.}
    \label{fig:Cattle2DCorrelationExample}
\end{figure}

The algorithm does not attempt to test specific scenarios.  However, we argue that the algorithm can be interpreted as quantifying the effect of the past scenarios that would have caused large losses for the current portfolio positions. \\
\\
As noted in Section \ref{sec:2DVisualization}, the algorithm is equivalent to counting the number of past historical days below and above an $K-1$ surface in $r_{k,j}/\sigma^{14}_{k,j-1}$ space that is orthogonal to the $<expo_{k,i-1} * \sigma^{14}_{k,i-1}>$ vector. The shape of the $K-1$ dimension surface changes according to the recent volatility and the current portfolio exposures.  The bigger the exposure and the higher the recent volatility of instrument $k$, the more important the direction is to the algorithm. \\
\\
Imagine the case where the current portfolio position is dominated by a spread on live cattle and feeder cattle futures after accounting for recent volatility (short feeder cattle, long live cattle). Feeder cattle are juvenile calves that mature into live cattle after finishing in feedlots.  The portfolio total return PDF can be visualized in Figure \ref{fig:Cattle2DCorrelationExample}. The lower-right side quadrant leads to losses. \\
\\
The precise locations of the dots far away from the line (in blue) are irrelevant to the location of the line. If these blue dots were to shift moderately, the line would not move because the worst 1 percent of days (in red) had not changed. \\
\\
However, locations of the dots close to the line (in orange) and below the line (in red) are of great importance.  Changes in the locations will lead to changes in the line's distance from the origin, which ultimately determines the predicted VaR. \\
\\
The orange and red days $j$ are not random.  They may be due to particular economic or political causes. Hypothetically (this example is purely hypothetical), let's suppose past orange and red days $j$ are primarily due to a sudden decrease in livestock feed prices leading to decreases in live cattle production cost, but increasing the feeder cattle value. If this were the case, the algorithm would effectively quantify how sudden livestock feed price changes affect the live vs feeder cattle spreads. In this way, we can interpret the algorithm as quantifying the effect of whatever scenarios that caused similar large losses in the past for the current portfolio positions, even if the algorithm does not attempt to articulate those causes in words. \\
\\
Crucially, the algorithm automatically adjusts which days $j$ (and the past causes of $j$) are considered important based on the current portfolio exposures and current recent volatilities.  Again, such adjustments directly incorporate the complex relationships between instruments in high-dimensional space, which may be difficult for humans to identify and interpret. \\
\\
This interpretation makes the algorithm complementary to algorithms such as \cite{Bouchaud99} that algorithmically identify the dominant factors most important to the current situation and then capture the non-linear, non-Gaussian effects of said factors during periods of stress.
\section{Back Testing with an Ensemble of 500 Portfolios with Random Positions}
\label{sec:EssembleTesting}
We back test our VaR algorithm using an ensemble of 500 portfolios.  Each portfolio trades all the following futures in Table \ref{tab:PortfolioFrontMonthFutures} and \ref{tab:PortfolioBackMonthFutures}. \\
\begin{table}[ht!]
    \centering
    \begin{tabular}{l l}
\rowcolor{gray!10}
        \bf Asset Class & \bf Futures Traded (Front Month) \\ 
        Volatility & VIX \\ 
\rowcolor{gray!10}
        Equity & S\&P 500, Nasdaq 100, Nikkei 225, FTSE 100, DAX \\ 
        Rates & Euribor \\
\rowcolor{gray!10}
        Gov. Bond & 10-Year Treasury, 5-Year Treasury, 2-Year Treasury, \\
\rowcolor{gray!10}
        & Euro-Bund, Long GILT \\ 
        Commodity Energy & WTI Crude, Natural Gas, Heating Oil \\ 
\rowcolor{gray!10}
        Commodity Metal & Copper, Gold, Palladium, Platinum, \\  
        Commodity Livestock & Lean Hog, Feeder Cattle, Live Cattle \\ 
\rowcolor{gray!10}
        Commodity Agriculture & Soybean, Soy Oil, Soy Meal,  \\
\rowcolor{gray!10}
        & Rough-Rice, Wheat, HRW Wheat, Corn \\ 
        Commodity Softs & Sugar, Coffee, Cotton, Cocoa \\ 
    \end{tabular}
    \caption{Traded front-month futures in test portfolios.}
    \label{tab:PortfolioFrontMonthFutures}
\end{table}
\begin{table}[ht!]
    \centering
    \begin{tabular}{l l}
\rowcolor{gray!10}
        \bf Asset Class & \bf Futures Traded (Back Month) \\ 
        Volatility & VIX \\ 
\rowcolor{gray!10}
        Commodity Energy & WTI Crude, Natural Gas, Heating Oil \\ 
        Commodity Metal & Copper, Gold, \\ 
\rowcolor{gray!10}
        Commodity Livestock & Lean Hog, Live Cattle \\ 
        Commodity Agriculture & Soybean, Soy Oil, Soy Meal,  \\
        & Rough-Rice, Wheat, HRW Wheat, Corn \\ 
\rowcolor{gray!10}
        Commodity Softs & Sugar \\ 
    \end{tabular}
    \caption{Traded back-month futures in test portfolios.}
    \label{tab:PortfolioBackMonthFutures}
\end{table}
\\
Each portfolio generates a random exposure for each instrument on for each day $i$ (front and the back month expiry are considered separate instruments with different weights) as defined in Equation \ref{eqn:randomExposure}.  No transaction costs are included because we are simply trying to test if the procedure can accurately predict the daily VaR with a large number of positions with complex relationships.  Obviously, no one will trade a randomly positioned portfolio. In reality, the VaR algorithm does not know the intended holding time of each position, so it cannot know the impact of trading. The user must have a separate method of quantifying expected slippage, liquidity constraints, and trading costs, all of which are outside the scope of this article. 
\begin{equation}
    expo_{k,i} = \frac{1}{1000} \frac{1}{\sigma^{252}_{k,i-1}} * RandomUniform(0.3,1.0) * RandomSign(-1,1)
    \label{eqn:randomExposure}
\end{equation}
The $\frac{1}{\sigma^{252}_{k,i-1}}$ component is the inverse average volatility in the past year up to day $i-1$.  $RandomUniform(0.3,1.0)$ is a uniform random value between 0.3-1.0, and $RandomSign(-1,1)$ is either -1 or 1 (long or short). \\
\\
Long-term volatility scaling is necessary because some instruments, such as natural gas, are orders of magnitude more volatile per unit exposure than others, such as US Treasuries. Without any volatility scaling, the portfolio would be equivalent to holding only a few of the most volatile instruments and other instruments will never be relevant to the portfolio's total return or risk.  \\
\\
The goal of the exercise is to test the performance of the algorithm at high dimensions (in this case, 49 dimensions).  We therefore need exposures where all tested assets are relevant for the total portfolio, instead of a few instruments always dominating. 
\section{Ensemble of Portfolio Back Test Results}
\label{sec:EssembleTestingResults}

\begin{figure}[ht!]
    \centering
    \includegraphics[width=0.75\linewidth]{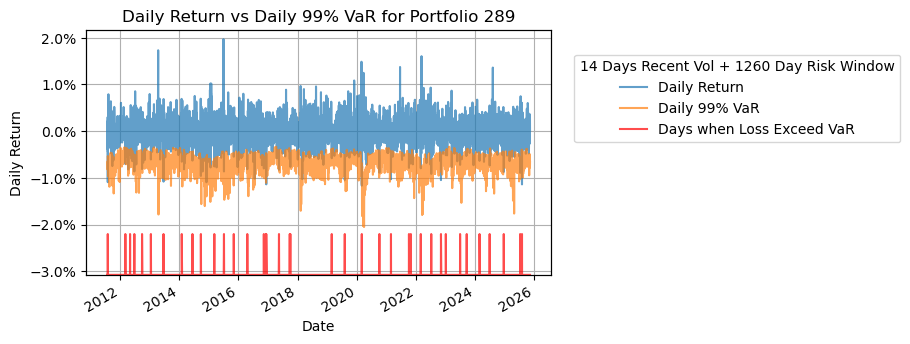}
    \caption{Example portfolio's daily return vs 99\% VaR.}
    \label{fig:portfolioReturnVsRisk}
\end{figure}

\begin{figure}[ht!]
    \centering
    \includegraphics[width=0.45\linewidth]{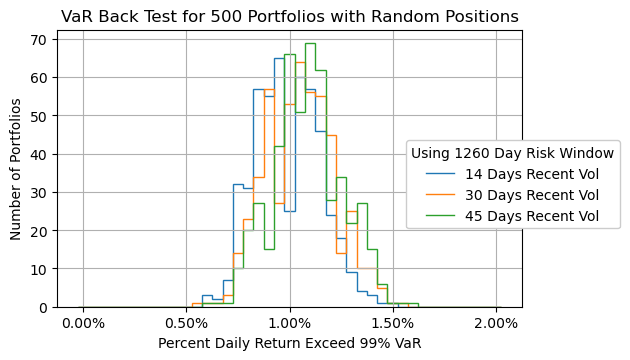}
    \includegraphics[width=0.45\linewidth]{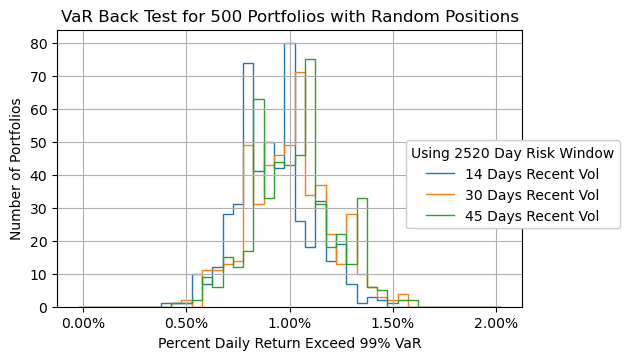}
    \includegraphics[width=0.45\linewidth]{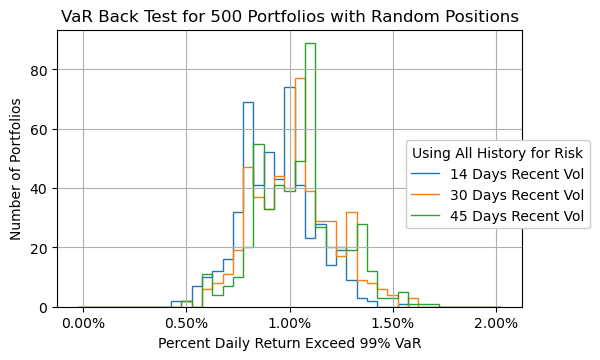}
    \caption{Rate of total portfolio loss exceeding the 99\% VaR estimate for different input parameters.}
    \label{fig:ensemblePortfolioLossExceedRiskRate}
\end{figure}

\begin{figure}[ht!]
    \centering
    \includegraphics[width=0.95\linewidth]{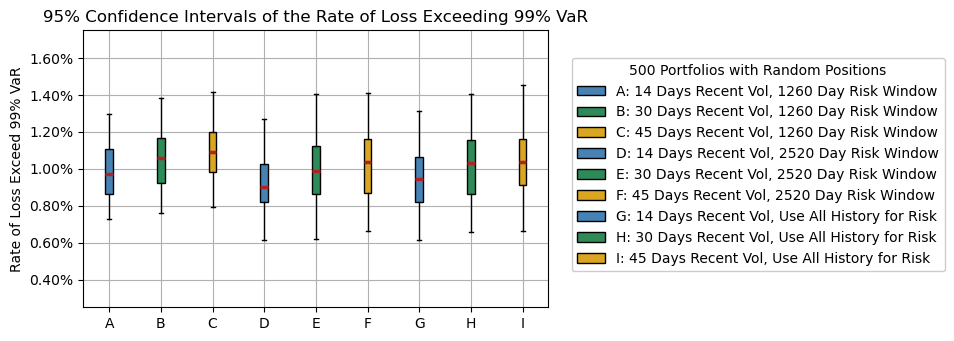}
    \caption{Box plot of the rate of loss exceeding 99\% VaR for different input parameters.}
    \label{fig:ensemblePortfolioLossExceedRiskRateBoxPlot}
\end{figure}

\begin{figure}[ht!]
    \centering
    \includegraphics[width=0.75\linewidth]{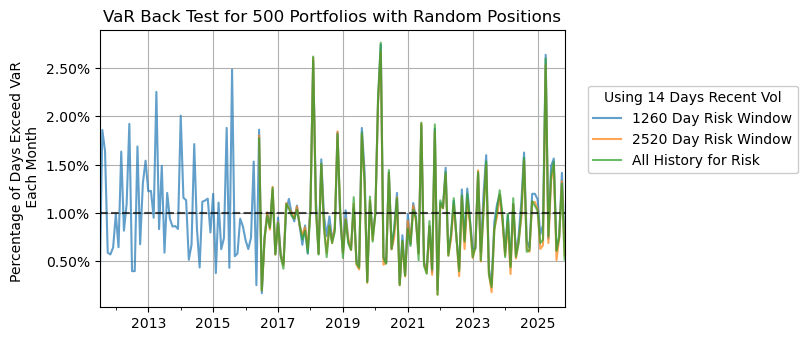}
    \caption{Month-by-month rate of loss exceeding estimated VaR vs time for 14-day recent volatility.}
    \label{fig:ensemblePortfolioLossExceedRiskVsTime_14DayRecentVol}
\end{figure}

\begin{figure}[ht!]
    \centering
    \includegraphics[width=0.45\linewidth]{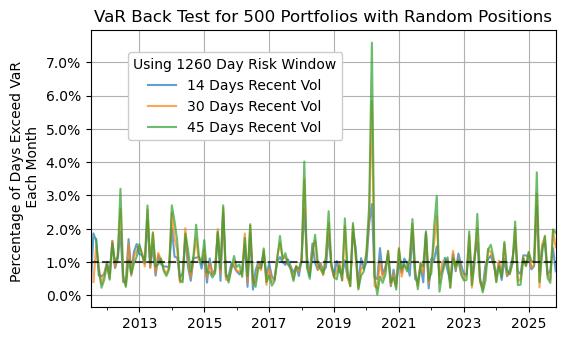}
    \includegraphics[width=0.45\linewidth]{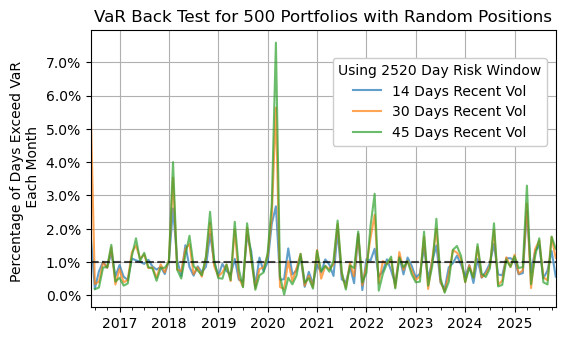}
    \includegraphics[width=0.45\linewidth]{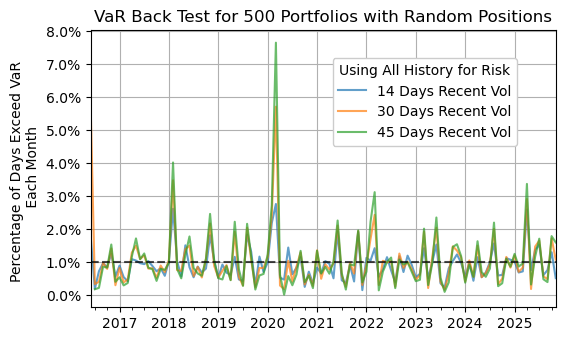}
    \caption{Month-by-month rate of loss exceeding estimated VaR vs time for $n =$ 14 days, 30 days, and 45 days recent volatility.}
    \label{fig:ensemblePortfolioLossExceedRiskVsTime_All}
\end{figure}

We compute the 99\% daily VaR estimator for each day $i$ using the exposures of each instrument on day $i-1$ and each instrument's historical data up to day $i-1$ using Equation \ref{eqn:portfolioTomorrowReturnPDF}. We also compute the portfolio's total return on day $i$ according to Equation \ref{eqn:portfolioTotalReturn}.\\
\begin{equation}
    total~portfolio~r_i = \sum_k expo_{k,i-1} * r_{k,i}
    \label{eqn:portfolioTotalReturn}
\end{equation}
We compare the portfolio's daily return and the 99\% daily VaR estimation and identify the days when the portfolio loss exceeds the VaR estimation as shown in Figure \ref{fig:portfolioReturnVsRisk}. The blue represent the portfolio's daily total return. The orange is the 99\% daily VaR estimated using the procedure in Section \ref{sec:PDF}.  The red tags the days when the portfolio return is more negative than the estimated 99\% daily VaR.\\
\\
We perform the above daily return vs VaR estimate calculation for 9 different combinations of VaR estimation input parameters: 3 different recent volatility lengths ($n=14$ days, 30 days, and 45 days for both $\sigma^{n}_{k,i-1}$ and $\sigma^n_{k,j-1}$) and 3 different historical $r_{k,j}/\sigma^n_{k,j-1}$ look back window lengths ($j$ $\in$ [the past 1260 days], [the past 2520 days], and [use all available history with a minimum of 2520 days before day $i$]). These two are the only manually set input parameters to the algorithm. \\
\\
The rate at which the portfolio's daily return is below the 99\% VaR estimate is shown in Figure \ref{fig:ensemblePortfolioLossExceedRiskRate}. By definition, 1 percent of all days should experience a portfolio total loss that exceeds the 99\% VaR if the VaR estimator is accurate.\\
\\
The median rate of loss exceeding the 99\% VaR estimate is between 0.90\% and 1.10\% for all combinations of VaR input parameters as seen in Figure \ref{fig:ensemblePortfolioLossExceedRiskRateBoxPlot} and Table \ref{tab:rateOfLoss}. The 68\% confidence intervals are within $1.0\pm0.3$\%, and the 95\% confidence intervals are within $1.0\pm0.5$\%. The whiskers in the box plot are at 2.5\% and 97.5\% quantiles. The red dash is the median, and the box represents the second and third quartiles. The ``Use All History" risk windows require a minimum of 2520 days before the first estimation of risk. \\
\begin{table}[ht!]
    \centering
    \begin{tabular}{c c c c c c} 
    \rowcolor{gray!20}
    \bf Recent Vol & \bf Risk Window & \bf 95\% Confidence & \bf 68\% Confidence & \bf 50\% Confidence & \bf Median \\
    14 Days & 1260 Days & 0.73\% - 1.30\% & 0.84\% - 1.16\% & 0.86\% - 1.11\% & 0.97\% \\
    \rowcolor{gray!20}
    30 Days & 1260 Days & 0.76\% - 1.38\% & 0.90\% - 1.22\% & 0.92\% - 1.17\% & 1.06\% \\
    45 Days & 1260 Days & 0.78\% - 1.42\% & 0.93\% - 1.25\% & 0.98\% - 1.20\% & 1.09\% \\
    \rowcolor{gray!20}
    14 Days & 2520 Days & 0.59\% - 1.29\% & 0.74\% - 1.11\% & 0.82\% - 1.02\% & 0.90\% \\
    30 Days & 2520 Days & 0.62\% - 1.40\% & 0.78\% - 1.20\% & 0.87\% - 1.12\% & 0.99\% \\
    \rowcolor{gray!20}
    45 Days & 2520 Days & 0.64\% - 1.41\% & 0.83\% - 1.24\% & 0.87\% - 1.16\% & 1.04\% \\
    14 Days & All History & 0.61\% - 1.31\% & 0.74\% - 1.11\% & 0.82\% - 1.06\% & 0.94\% \\
    \rowcolor{gray!20}
    30 Days & All History & 0.66\% - 1.42\% & 0.82\% - 1.20\% & 0.87\% - 1.15\% & 1.03\% \\
    45 Days & All History & 0.64\% - 1.45\% & 0.83\% - 1.24\% & 0.91\% - 1.16\% & 1.04\% \\
    \end{tabular}
    \caption{Rate of loss exceeding the estimated 99\% VaR for an ensemble of 500 portfolios.}
    \label{tab:rateOfLoss}
\end{table}
\\
The algorithm is able to accurately estimate 99\% VaR for a portfolio with 49 instruments, with the median portfolio's daily returns breaking through the VaR estimation at $1.0\pm0.1$\% of all days. \\
\\
The rate of loss exceeding 99\% VaR estimation vs time, averaged over each month, is shown in Figure \ref{fig:ensemblePortfolioLossExceedRiskVsTime_14DayRecentVol}. A 14-day recent volatility is used for VaR calculation with 3 different $r_{k,j}/\sigma^{14}_{k,j-1}$ look-back window lengths. \\
\\
Here, we use the 14-day recent volatility for the VaR calculation. We see that regardless of the $r_{k,j}/\sigma^{14}_{k,j-1}$ look-back window length we get a roughly steady rate of loss each month centered around 1\%. \\
\\
The monthly average of the rate of loss exceeding VaR can increase to 2.5\% to 3.0\% during certain, especially volatile months, such as March 2020.  However, each month only contains around 22 trading days, so 3\% of 22 days is only 0.66 days out of the month.  \\
\\
In other words, we do not see a large pile-up of days where losses exceed the estimated VaR during especially volatile times.  This, in turn, means the estimated VaR is able to forecast when the portfolio positions are low-risk and when the portfolio positions are high-risk by rapidly adapting to the current situation. \\
\\
The same plot is made for longer 30-day and 45-day recent volatility averages in Figure \ref{fig:ensemblePortfolioLossExceedRiskVsTime_All}. Here, we can see significantly higher spikes (6 and 8 percent) rate of loss exceeding the VaR estimations during events such as March 2020 compared to the 14-day recent volatility results (spikes to 3 percent) in Figure \ref{fig:ensemblePortfolioLossExceedRiskVsTime_14DayRecentVol}. \\
\\
We see the importance of the fast adaptability of the 14-day recent volatility. The short number of days used in the root-mean-square computation means that the $\sigma^{14}_{k,i-1}$ rapidly adjusts to changes in recent conditions.  \\
\\
Volatility expansion occurs during periods of acute distress, such as March 2020.  However, this week's daily true range percentages will be included in next week's $\sigma^{14}_{k,i-1}$.  If the 1 percent worst loss tail is estimated to be at 4x the recent $\sigma^{14}_{k,i-1}$ core volatility, then the volatility must expand geometrically week after week for us to get sustained break-throughs of our VaR estimates over time.  This level of geometric expansion is difficult, if not impossible, to sustain for more than 1-2 weeks maximum.  Very quickly, $4*4*4=64$, and the daily volatility would need to increase beyond 100\% for such a rate of geometric expansion.\\
\\
In other words, the fast 14-day recent volatility estimate gives a balance between a stable, statistically significant RMS average, all the while not sacrificing fast adaptability to recent conditions. 
\section{Effect of an Additional Day Delay}

\begin{figure}[ht!]
    \centering
    \includegraphics[width=0.95\linewidth]{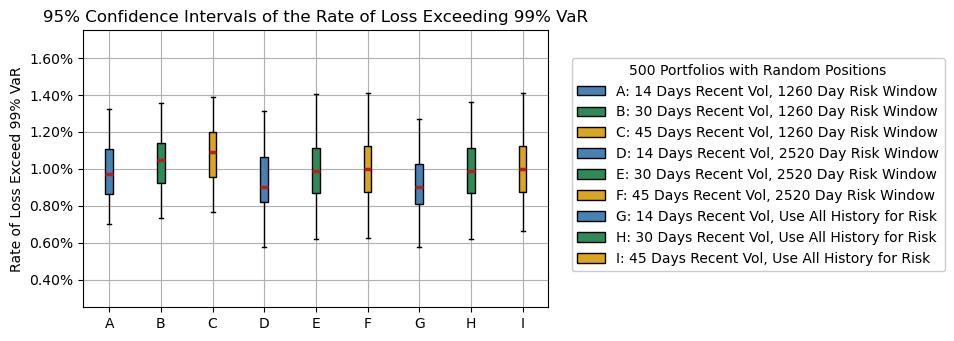}
    \caption{Box plot of the rate of loss exceeding 99\% VaR for different VaR input parameters when the VaR for day $i$ is estimated without knowledge of day $i-1$'s performance data.}
    \label{fig:ensemblePortfolioLossExceedRiskRateBoxPlot1AdditionalDayDelay}
\end{figure}

\begin{figure}[ht!]
    \centering
    \includegraphics[width=0.75\linewidth]{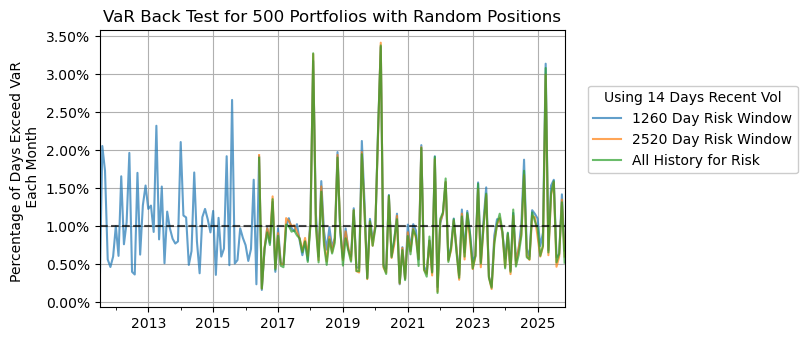}
    \caption{Month-by-month rate of loss exceeding estimated VaR vs time when the VaR for day $i$ is estimated without knowledge of day $i-1$'s performance data.}
    \label{fig:ensemblePortfolioLossExceedRiskVsTime_14DayRecentVol1AdditionalDayDelay}
\end{figure}

We present the situation where we are currently in the middle of trading day $i-1$.  None of the assets have closed for the day, so we do not know the high, low, and close of the day $i-1$.  We do know all the positions ($expo_{k,i-1}$) we would like to take at the close of $i-1$. We would like to predict the VaR of future day $i$ of a portfolio with $expo_{k,i-1}$ before committing to the position.  In other words, we use trading data up to day $i-2$ and $expo_{k,i-1}$ to predict the portfolio VaR on $i$. \\
\\
In this case, we add an additional 1-day delay to Equation \ref{eqn:portfolioTomorrowReturnPDF} to get Equation \ref{eqn:portfolioTomorrowReturnPDF1DayDelay}. \\
\begin{equation}
    \centering
    \probP(portfolio~return_i) = \sum_k expo_{k,i-1} * \sigma^{14}_{k,i-2} * \frac{r_{k,j}}{\sigma^{14}_{k,j-2}}
    \label{eqn:portfolioTomorrowReturnPDF1DayDelay}
\end{equation}
The recent volatility $\sigma^{14}_{k,i-2}$ is comprised of data from day $i-15$ to $i-2$.  The historical ratio $r_{k,j}/\sigma^{14}_{k,j-2}$ is the ratio between the return on the past day $j$ and its own $\sigma^{14}_{k,j-2}$ which is comprised of data from day $j-15$ to $j-2$. The historical days $j$ include the set of days up to $i-2$. \\
\\
We get similar results for the rate of loss exceeding the 99\% VaR estimate with the additional day delay in Figure \ref{fig:ensemblePortfolioLossExceedRiskRateBoxPlot1AdditionalDayDelay} and Table \ref{tab:rateOfLoss1AdditionalDayDelay}.  Medians are within $1.0\pm0.1$\%.  50\% confidence are within $1.0\pm0.2$\%. 68\% confidence are within $1.0\pm0.3$\% and 95\% confidence are within $1.0\pm0.5$\%.\\
\\
Note, all portfolios' instrument exposures are identical to the portfolios with no additional delay.  Only the VaR calculation is different. \\
\begin{table}[ht!]
    \centering
    \begin{tabular}{c c c c c c} 
    \rowcolor{gray!20}
    \bf Recent Vol & \bf Risk Window & \bf 95\% Confidence & \bf 68\% Confidence & \bf 50\% Confidence & \bf Median \\
    14 Days & 1260 Days & 0.70\% - 1.32\% & 0.84\% - 1.16\% & 0.86\% - 1.11\% & 0.97\% \\
    \rowcolor{gray!20}
    30 Days & 1260 Days & 0.73\% - 1.36\% & 0.87\% - 1.22\% & 0.92\% - 1.14\% & 1.04\% \\
    45 Days & 1260 Days & 0.76\% - 1.39\% & 0.90\% - 1.25\% & 0.95\% - 1.20\% & 1.09\% \\
    \rowcolor{gray!20}
    14 Days & 2520 Days & 0.57\% - 1.31\% & 0.74\% - 1.11\% & 0.82\% - 1.07\% & 0.90\% \\
    30 Days & 2520 Days & 0.62\% - 1.40\% & 0.78\% - 1.20\% & 0.87\% - 1.11\% & 0.99\% \\
    \rowcolor{gray!20}
    45 Days & 2520 Days & 0.62\% - 1.41\% & 0.83\% - 1.20\% & 0.87\% - 1.12\% & 1.00\% \\
    14 Days & All History & 0.57\% - 1.27\% & 0.74\% - 1.11\% & 0.81\% - 1.02\% & 0.90\% \\
    \rowcolor{gray!20}
    30 Days & All History & 0.62\% - 1.36\% & 0.78\% - 1.20\% & 0.87\% - 1.11\% & 0.99\% \\
    45 Days & All History & 0.64\% - 1.41\% & 0.83\% - 1.20\% & 0.87\% - 1.12\% & 1.00\% \\
    \end{tabular}
    \caption{Rate of loss exceeding the estimated 99\% VaR when the VaR for day $i$ is estimated without knowledge of day $i-1$'s performance data.  }
    \label{tab:rateOfLoss1AdditionalDayDelay}
\end{table}
\\
The rate of loss exceeding VaR estimation vs time for 14-day recent volatility now spikes to between $3.0$\% to $3.5$\% during especially volatile months (compared to $2.5-3.0$\% without additional delay), as shown in Figure \ref{fig:ensemblePortfolioLossExceedRiskVsTime_14DayRecentVol1AdditionalDayDelay}. The additional delay slows down the reaction time of the recent volatility $\sigma^{14}_{k,i-1}$, lowering its adaptability to the current situation. \\
\\
The effect of having an additional day delay essentially slows the adaptability of the estimator to recent conditions.  When volatility increases, the algorithm is slightly slower to adapt and risk estimation is lower compared to if there were no delay. In this situation, we should expect to see more breakthroughs of the VaR estimation.  Likewise, when volatility decreases, the algorithm is also slower to adapt, and we see fewer breakthroughs of the VaR estimation compared to if there were no delay.  In the long term, the median rate of loss exceeding VaR estimation is consistent with those estimated with no additional delay.
\section{Benefits and Limitations of the Algorithm}

Perhaps the biggest benefit is that the algorithm requires no manual analysis of the specific asset classes in the portfolio.  The only required input is the past historical daily candles of the instruments.  The algorithm directly takes into account all the relationships between the assets, and the non-linear skew in past data.  VIX futures, a highly right-skewed and non-linear instrument, with a strong and complex relationship to equities, and significant relationships to energy, gov. bonds, rates, and metals, are incorporated in the example test.  This practical benefit makes the algorithm useful as a baseline case when specific dynamics of an asset class or the relationship between an asset class and other asset classes are difficult to manually disentangle. \\
\\
The algorithm is computationally light and can be run in real time.  Computationally, the algorithm consists of only vector additions for K instruments and computing a quantile. \\
\\
Another benefit is that the algorithm works on underlying positions instead of the time series of the total portfolio.  The algorithm will return the same risk for a certain exposure regardless of how the decision is made to target such an exposure.  \\
\\
Risk algorithms that take the time series of the past portfolio or trading strategy performance as an input run the risk of computing risk according to the past selection power of the trading strategy. Philosophically speaking, the purpose of risk management is to ask ``what would happen if the trading strategy is wrong?"  It is therefore dangerous to answer this question according to how successful the strategy was in the past.  \\
\\
Analyzing risk based on past behavior of the position alone, without relying on the trader's selection power, is the algorithmic embodiment of the physical separation of the risk management and alpha-generating departments within an organization.  All risk managers should be skeptical if a trader approaches with the argument ``Trust me, I was right in the past, and so my current positions are not risky." Likewise, risk estimation algorithms should be skeptical of past performance (or worse, back-tested past performance) as an indicator of low future risk. Allocation algorithms are free to separately take the expected returns as an input parameter if desired. \\
\\
The clearest limitation is that the algorithm requires all underlying instruments to have past data day $j$ in order to work.  We are therefore limited by the instrument with the shortest amount of past data.  In our example, VIX futures, which only began trading in 2004, forms our limit.  Back-month expiry VIX futures only have consistent trading volumes since 2006. We then need another 1260 trading days (depending on the historical data window) of data before we can estimate the first VaR. \\
\\
Instruments with short historical data can be back-filled using simulations and mixed with actual historical data from instruments that have long historical data as shown in Table \ref{tab:Simulation}.  The simulation models how the instrument would have behaved on specific past historical days $j$ for which no actual data exists.  In this way, we use as much historical data as possible while being clear about what data is historical and what is based on simulation. \\
\begin{table}[ht!]
    \centering
    \begin{tabular}{c c}
    \rowcolor{gray!20}
       \bf Instrument  & \bf Available Historical (\swl{-}) and Simulation (\swl{S}) Data \\
       1  & 
       \swl{S}\swl{S}\swl{S}\swl{S}\swl{S}\swl{S}\swl{-}\swl{-}\swl{-}\swl{-}\swl{-}\swl{-}\swl{-}\swl{-}\swl{-}\swl{-}\swl{-}\swl{-}\swl{-}\swl{-}\swl{-}\swl{-}\swl{-}\swl{-}\swl{-}\swl{-}\swl{-}\swl{-}\swl{-}\swl{-}\swl{-}\swl{-}\swl{-}\swl{-}\swl{-}\swl{-}\swl{-}\swl{-}\swl{-}\swl{-}\swl{-}\swl{-}\swl{-}\swl{-}\swl{-}\swl{-}\swl{-}\swl{-} \\
    \rowcolor{gray!20}
       2  & 
       \swl{-}\swl{-}\swl{-}\swl{-}\swl{-}\swl{-}\swl{-}\swl{-}\swl{-}\swl{-}\swl{-}\swl{-}\swl{-}\swl{-}\swl{-}\swl{-}\swl{-}\swl{-}\swl{-}\swl{-}\swl{-}\swl{-}\swl{-}\swl{-}\swl{-}\swl{-}\swl{-}\swl{-}\swl{-}\swl{-}\swl{-}\swl{-}\swl{-}\swl{-}\swl{-}\swl{-}\swl{-}\swl{-}\swl{-}\swl{-}\swl{-}\swl{-}\swl{-}\swl{-}\swl{-}\swl{-}\swl{-}\swl{-} \\
       3  & 
       \swl{S}\swl{S}\swl{S}\swl{S}\swl{S}\swl{S}\swl{S}\swl{S}\swl{S}\swl{S}\swl{S}\swl{S}\swl{S}\swl{S}\swl{-}\swl{-}\swl{-}\swl{-}\swl{-}\swl{-}\swl{-}\swl{-}\swl{-}\swl{-}\swl{-}\swl{-}\swl{-}\swl{-}\swl{-}\swl{-}\swl{-}\swl{-}\swl{-}\swl{-}\swl{-}\swl{-}\swl{-}\swl{-}\swl{-}\swl{-}\swl{-}\swl{-}\swl{-}\swl{-}\swl{-}\swl{-}\swl{-}\swl{-} \\
    \rowcolor{gray!20}
       4  & 
       \swl{-}\swl{-}\swl{-}\swl{-}\swl{-}\swl{-}\swl{-}\swl{-}\swl{-}\swl{-}\swl{-}\swl{-}\swl{-}\swl{-}\swl{-}\swl{-}\swl{-}\swl{-}\swl{-}\swl{-}\swl{-}\swl{-}\swl{-}\swl{-}\swl{-}\swl{-}\swl{-}\swl{-}\swl{-}\swl{-}\swl{-}\swl{-}\swl{-}\swl{-}\swl{-}\swl{-}\swl{-}\swl{-}\swl{-}\swl{-}\swl{-}\swl{-}\swl{-}\swl{-}\swl{-}\swl{-}\swl{-}\swl{-} \\
       5  & 
       \swl{S}\swl{S}\swl{S}\swl{S}\swl{S}\swl{S}\swl{S}\swl{S}\swl{S}\swl{S}\swl{S}\swl{S}\swl{S}\swl{S}\swl{S}\swl{S}\swl{S}\swl{S}\swl{S}\swl{S}\swl{S}\swl{S}\swl{S}\swl{S}\swl{S}\swl{S}\swl{S}\swl{S}\swl{S}\swl{S}\swl{S}\swl{S}\swl{S}\swl{S}\swl{-}\swl{-}\swl{-}\swl{-}\swl{-}\swl{-}\swl{-}\swl{-}\swl{-}\swl{-}\swl{-}\swl{-}\swl{-}\swl{-} \\
    \rowcolor{gray!20}
       6  & 
       \swl{-}\swl{-}\swl{-}\swl{-}\swl{-}\swl{-}\swl{-}\swl{-}\swl{-}\swl{-}\swl{-}\swl{-}\swl{-}\swl{-}\swl{-}\swl{-}\swl{-}\swl{-}\swl{-}\swl{-}\swl{-}\swl{-}\swl{-}\swl{-}\swl{-}\swl{-}\swl{-}\swl{-}\swl{-}\swl{-}\swl{-}\swl{-}\swl{-}\swl{-}\swl{-}\swl{-}\swl{-}\swl{-}\swl{-}\swl{-}\swl{-}\swl{-}\swl{-}\swl{-}\swl{-}\swl{-}\swl{-}\swl{-} \\
       7  & 
       \swl{S}\swl{S}\swl{S}\swl{-}\swl{-}\swl{-}\swl{-}\swl{-}\swl{-}\swl{-}\swl{-}\swl{-}\swl{-}\swl{-}\swl{-}\swl{-}\swl{-}\swl{-}\swl{-}\swl{-}\swl{-}\swl{-}\swl{-}\swl{-}\swl{-}\swl{-}\swl{-}\swl{-}\swl{-}\swl{-}\swl{-}\swl{-}\swl{-}\swl{-}\swl{-}\swl{-}\swl{-}\swl{-}\swl{-}\swl{-}\swl{-}\swl{-}\swl{-}\swl{-}\swl{-}\swl{-}\swl{-}\swl{-} \\
    \rowcolor{gray!20}
       8  & 
       \swl{S}\swl{S}\swl{S}\swl{S}\swl{S}\swl{-}\swl{-}\swl{-}\swl{-}\swl{-}\swl{-}\swl{-}\swl{-}\swl{-}\swl{-}\swl{-}\swl{-}\swl{-}\swl{-}\swl{-}\swl{-}\swl{-}\swl{-}\swl{-}\swl{-}\swl{-}\swl{-}\swl{-}\swl{-}\swl{-}\swl{-}\swl{-}\swl{-}\swl{-}\swl{-}\swl{-}\swl{-}\swl{-}\swl{-}\swl{-}\swl{-}\swl{-}\swl{-}\swl{-}\swl{-}\swl{-}\swl{-}\swl{-} \\
    \end{tabular}
    \caption{Hypothetical example of incorporating single instrument simulations.}
    \label{tab:Simulation}
\end{table}
\\
Another method for mitigating the lack of historical data is to simulate 95\% or 90\% confidence VaR instead of 99\%.  The limit for the algorithm is essentially the need for a statistically significant number of past days below the integration surface.  For the same set of historical data, 95\% VaR has 5 times the amount of data below the integration surface as 99\% VaR. \\
\\
We would argue that if we only have available 200 days of data, with no additional simulation based on similar assets with longer histories, then we are inherently limited in our ability to estimate the 99\% VaR no matter the computation method.  There simply does not exist enough data to say anything statistically significant about an event that is expected to have occurred only 2 times in the past. \\
\\
Another limitation is that the algorithm does not capture the autocorrelation between past days $j$.  The recent volatility $\sigma^{n}_{k,i-1}$ is inherently an auto-correlated quantity. However, the algorithm treats the historical ratio $r_{k,j}/\sigma^{n}_{k,j-1}$ as independent points and ignores the relationships between past days $j$.  In practice, we see that the algorithm's rate of portfolio level loss exceeding estimated VaR is relatively stable with time in Figure \ref{fig:ensemblePortfolioLossExceedRiskVsTime_14DayRecentVol} and \ref{fig:ensemblePortfolioLossExceedRiskVsTime_14DayRecentVol1AdditionalDayDelay} as long as the $n$ days recent volatility is short (14 days).  Therefore, the algorithm remains accurate even in past instances of extreme volatility, where we expect autocorrelation effects such as volatility expansion to be the strongest.
\section{Conclusion}
We presented a computationally simple, generic value-at-risk algorithm that works for an arbitrarily high number of dimensions. \\
\\
The algorithm breaks down the probability density function of the portfolio's total return into 3 components:
\begin{itemize}
    \item $expo_{k,i-1}$: the exposure to financial instrument $k$ as measured by market value / AUM at the close of day $i-1$. 
    \item $\sigma^{14}_{k, i-1}$: the recent volatility of $k$ in the most recent 14 days before day $i$.
    \item $\frac{r_{k,j}}{\sigma^{14}_{k,j-1}}$: how much $k$'s return on each past day $j$ deviated from the recent volatility before day $j$.
\end{itemize}
The portfolio total return PDF can then be used to calculate the VaR at any confidence, and the expected shortfall of the user's choice. \\
\\
The recent volatility $\sigma^{14}_{k, i-1}$ rapidly adapts to the recent condition. The historical $r_{k,j}/\sigma^{14}_{k,j-1}$ ratio captures the non-Gaussian skew within a single instrument and all relations in the past between different instruments $k$. \\
\\
The algorithm is mathematically equivalent to Monte-Carlo integration above and below a $K-1$ dimensional surface in the $K$ dimensional $r_{k,j}/\sigma^{14}_{k,j-1}$ space, where each past day $j$ is a Monte Carlo trial as discussed in Section \ref{sec:2DVisualization}.  Monte-Carlo integration does not lose accuracy at high dimensions and can directly incorporate all $ K$-dimensional relationships simultaneously. The algorithm applies math that works in high dimensions directly to high-dimensional historical data with no assumptions, approximations, or compression. \\
\\
The algorithm is computationally light and needs no a priori knowledge of the underlying asset classes.  Hence, the algorithm can serve as a benchmark VaR estimator in the absence of any manual analysis of the specific assets or the complex relationships between assets in high-dimensional space.  \\
\\
We tested the algorithm with an ensemble of 500 portfolios with daily random positions in 49 different liquid futures, expiration combinations (long/short, front+back expiry, VIX, equity indexes, government bonds, interest rates, energy, metals, livestock, agriculture, and softs) in section \ref{sec:EssembleTesting} and \ref{sec:EssembleTestingResults}.  All VaR estimation is made blind to the future at all points in time. \\
\\
The algorithm is accurate for this high number of instruments with complex relationships. Median portfolio's rate of loss exceeding the 99\% VaR estimate is between $1.0\pm0.1$\% for all combinations of tested input parameters. By definition, 1\% of daily returns should exceed the loss predicted by the 99\% daily VaR if the VaR estimator is accurate.  The 68\% confidence of the portfolio's rate of loss exceeding the 99\% VaR estimate is between $1.0\pm0.3\%$. The 95\% confidence of the portfolio's rate of loss exceeding the 99\% VaR estimate is between $1.0\pm0.5\%$. \\
\\
The 14-day recent volatility adapts fast enough that even in the worst months of market volatility, the rate of portfolio total return breaking through the estimated VaR level remains stable. The rate of loss exceeding 99\% VaR estimations spikes to 3\% during exceptionally volatile months such as COVID March 2020 and Volmageddon in 2018.  In perspective, 3\% in one month corresponds roughly to 0.66 days per portfolio with a loss exceeding the 99\% VaR estimation. Hence, no significant pile-up is observed at any point in time. \\
\\
The algorithm can compute VaR or CVaR at any confidence, as the algorithm computes the total portfolio return PDF.  We leave back-testing of CVaR estimates as a future exercise.  Weekly, bi-weekly, or monthly VaR variations of the algorithm are also possible with modifications, but again are beyond the scope of this article.

\clearpage

\section{Declaration of Interest}
The author reports no conflicts of interest. The author alone is responsible for the content and writing of the paper.

\section{Acknowledgements}

Thank you, Alberto Desirelli (CERN Pension Fund) and Jean-Philippe Bouchaud (CFM), for your help editing, suggestions, and guidance through the publication process.  Thank you, Sandrine Ungari (Société Générale) and Brian Fleming (Société Générale), for the thoughtful discussions and for your help with editing. Thank you, Jerome Callut, Jeremie Canoni-Meynet, and Anthony Dearden of DCM Systematic for independently testing the algorithm on DCM’s infrastructure.  Thank you, Matteo Erigozzi and Roberto Segala, for your work on the code and data infrastructure used at Investivity. Thank you, Pierluigi Catastini (Investivity), for giving support and academic freedom in pursuing publication. 

\bibliographystyle{apacite}
\bibliography{sample.bib}

@article{HongMCSummary,
author = {Hong, L. Jeff and Hu, Zhaolin and Liu, Guangwu},
title = {Monte Carlo Methods for Value-at-Risk and Conditional Value-at-Risk: A Review},
year = {2014},
issue_date = {August 2014},
publisher = {Association for Computing Machinery},
address = {New York, NY, USA},
volume = {24},
number = {4},
issn = {1049-3301},
url = {https://doi.org/10.1145/2661631},
doi = {10.1145/2661631},
journal = {ACM Trans. Model. Comput. Simul.},
month = nov,
articleno = {22},
numpages = {37},
}

@book{Bouchaud99,
    author = "Bouchaud, Jean-Philippe and Potters, Marc",
    title = "Theory of Financial Risk and Derivative Pricing",
    publisher = "Cambridge University Press",
    year = "1999",
    chapter = "Worst Fluctuation Method for Fast Value-at-Risk Estimates",
    pages = "220--223",
    ISBN =  {9780521741866}
}

@article{BouyéCopula,
    author = "Bouyé, Eric and Salmon, Mark Howard",
    title = "Measuring the Dependence between Non-Gaussian Financial Assets Using Copulae: Risk Management, Option Pricing and Default Risk ",
    journal = "Working Paper",
    year = "2008",
    url = "http://dx.doi.org/10.2139/ssrn.2708678"
}

@article{HongKernel,
    author  = "L. Jeff Hong and Sandeep Juneja and Guangwu Liu",
    title   = "Kernel Smoothing for Nested Estimation with Application to Portfolio Risk Measurement.",
    year    = "2017",
    journal = "Operations Research",
    volume  = "65",
    number  = "3",
    pages   = "657--673",
    url = "https://doi.org/10.1287/opre.2017.1591"
}

@article{LopezHRP,
    author  = "López de Prado, Marcos",
    title   = "Building Diversified Portfolios that Outperform Out-of-Sample",
    year    = "2016",
    journal = "The Journal of Portfolio Management",
    url = "https://doi.org/10.3905/jpm.2016.42.4.059",
    volume  = "42",
    number  = "4",
    pages   = "59--69"
}

@article{GlassermanIS,
 ISSN = {00251909, 15265501},
 URL = {http://www.jstor.org/stable/2661652},
 author = {Paul Glasserman and Philip Heidelberger and Perwez Shahabuddin},
 journal = {Management Science},
 number = {10},
 pages = {1349--1364},
 publisher = {INFORMS},
 title = {Variance Reduction Techniques for Estimating Value-at-Risk},
 urldate = {2025-11-20},
 volume = {46},
 year = {2000}
}

@ARTICLE{MeanVariance,
title = {PORTFOLIO SELECTION},
author = {Markowitz, Harry},
year = {1952},
journal = {Journal of Finance},
volume = {7},
number = {1},
pages = {77-91},
url = {https://EconPapers.repec.org/RePEc:bla:jfinan:v:7:y:1952:i:1:p:77-91}
}

@misc{bardou2010computationvarcvarusing,
      title={Computation of VaR and CVaR using stochastic approximations and unconstrained importance sampling}, 
      author={Olivier Aj Bardou and Noufel Frikha and G. Pagès},
      year={2010},
      eprint={0812.3381},
      archivePrefix={arXiv},
      primaryClass={q-fin.CP},
      url={https://arxiv.org/abs/0812.3381}, 
}

\end{document}